# Adsorption behavior of conjugated {C}$_3$-oligomers on Si(100) and HOPG surfaces

G. Mahieu, B. Grandidier*, D. Stiévenard, C. Krzeminski, C. Delerue

*Institut d'Electronique et de Microélectronique du Nord, IEMN, (CNRS, UMR 8520)*

*Département ISEN, 41 bd Vauban, 59046 Lille Cédex (France)*

C. Martineau, J. Roncali

*Ingénierie Moléculaire et Matériaux Organiques,*

*CNRS UMR 6501, Université d'Angers, 2 bd Lavoisier, 49045 Angers (France)*

A π-conjugated {C}$_{3h}$-oligomer involving three dithienylethylene branches bridged at the meta positions of a central benzenic core has been synthesized and deposited either on the Si(100) surface or on the HOPG surface. On the silicon surface, scanning tunneling microscopy allows the observation of isolated molecules. Conversely, by substituting the thiophene rings of the oligomers with alkyl chains, a spontaneous ordered film is observed on the HOPG surface. As the interaction of the oligomers is different with both surfaces, the utility of the Si(100) surface to characterize individual oligomers prior to their use into a 2D layer is discussed.

# Introduction

Two dimensional (2D) molecular arrangements of conjugated oligomers can lead to the formation of novel nanostructures, which could take benefit of the oligomer electronic properties. Different techniques exist to form such monolayers, like the vacuum deposition, Langmuir-Blodgett or self- assembly techniques. Among all those techniques, the easiest one consists of depositing a drop of solution containing oligomers on a substrate.[1,2] On atomically flat substrates, which interact weakly with the molecules, *via* van der Waals forces for example, the oligomers can form two dimensional layers at the solid-liquid interface. Using this simple deposition technique, most of the 2D molecular layer studies based on oligomers have been achieved with simple one dimensional oligomeric chain. Only recently, arrangements of more complex oligomers have been investigated.[3]

A large number of 2D molecular layers have been formed on highly pyrolytic graphite (HOPG), as such a system is well suited to scanning tunneling microscopic (STM) experiments in air. Although this technique has the potential to provide the arrangement of 2D layers, the resolution of the molecular features is generally poor. Indeed, the STM images suffer from the drift of the microscope and the instability of the tunneling junction due to the experimental conditions. As the oligomers deposited on HOPG become more and more complex, a good understanding of their arrangement requires the observation of isolated oligomers and the recognition of their subcomponents prior to the formation of a molecular layer. Such a condition can be difficult to obtain with metallic surfaces, since their interaction with the molecules is generally weak and the molecules easily diffuse at room temperature.[4,5,6] An alternative could be the use of semiconductor surfaces.

Here we report on the synthesis of an oligomer possessing a ternary symmetry. Prior to the deposition of the oligomer on HOPG to form a film, the oligomers are vapour deposited onto a silicon surface in ultra high vacuum (UHV). We show that such a surface allows the identification of the oligomer molecular structure at room temperature. By attaching alkyl chains to the oligomer backbone, the deposition of a drop containing those molecules leads to the formation of a 2D layer on HOPG. The knowledge of the molecular structures observed on the Si(100) surface gives strong support for an arrangement with a honeycomb structure on HOPG. While the Si(100) surface provides a good mean to characterize individual oligomers, the comparison of molecular features between both surfaces requires however some cautions, which are discussed.

## Experimental Section

**Oligomer synthesis.** As shown in figure 1, the target molecule **1** consists of three dithienylethylene conjugated branches attached at the meta positions of a central benzenic core through an ethylene linkage. As shown in previous works, oligothienylenevinylene oligomers present the smallest HOMO-LUMO gap among known conjugated oligomers.[7] This property results from the presence of ethylene linkages, which prevent rotational disorder and contribute to decrease the overall aromatic character thus allowing optimal π-electron delocalization along the branches. Tris-1,3,5-bromomethylbenzene **4** was prepared according to a known procedure.[8] This compound was then converted into the tris-phosphonate **3** by reaction with triethylphosphite (yield 95%). The target compound was then obtained in 45% yield by a triple Wittig reaction between compound **3** and the carboxaldehyde of dithienylethylene **2**. The characterization of the target compound by usual spectroscopic and analytical methods gave results in full agreement with the expected chemical structure.

To enhance the solubility of the molecule **1** for its deposition on HOPG, the aldehyde **2** was substituted by hexyl-chains at the β,β'-positions of each thiophene.[9] As the molecule **1** substituted by hexyl chains is made up of oligothienylenevinylenes (nTV),[7,10] it is noted {C}$_{3h}$-2TV in the following.

**1,3,5-tris-(diethoxyphosphinylmethyl)-benzene 3.** A mixture of 1,3,5 bromomethyl benzene **4** (4.24g 11.89 mmol) and triethylphophite (6.5 mL, 37.86 mmol) is refluxed for 3h. Evaporation of triethylphosphite gave 6g (95%) of an oil. $^1$H NMR (CDCl$_3$) 7.10 (d, 3H, $^3J$ = 2.3 Hz), 3.98 (quint, 12H, $^3J$ = 7.16 Hz), 3.08 (d, 6H, $^2J$ = 22.01 Hz), 1.23-1.2 (m, 18H).

**Tris-1,3,5-{(E)-1-[5-[(E)-(2-thiényl)ethen-1-yl]-2-thienyl]ethen-2-yl} benzene 1.** Triphosphonate **3** (1.02 g, 1.94 mmol) and aldehyde **2** (1.5 g, 6.8 mmol) are dissolved in 100 mL of anhydrous THF at 0°C under a nitrogen atmosphere. Potassium terbutylate (1.53 g, 13 mmol) is added portionwise and the mixture is stirred at room temperature for 15 h. After addition of methanol, the precipitate is filtered and recrystallized twice in chloroform to give 0.66 g (45%) of a brown solid. m.p. 208-212°C. ms (FAB+) m/z 726 (M$^{+\cdot}$ ; 100). $^1$H NMR (DMSO) : 7.69 (s, 3H, H2, H4, H6) ; 7.58 (d, 3H, $^3J$ = 16.05 Hz, H8) ; 7.48 (d, 3H, $^3J$ = 4.58 Hz, H18) ; 7.25 (d, 3H, $^3J$ = 3.52 Hz, H16) ; 7.17 (dd, 6H, $^3J$ = 3.76 Hz, H10, H11) ;7.14 (s, 6H, H13, H14) ; 7.07 (dd, 3H, $^3J$ = 4.93 Hz, $^3J$ = 3.52 Hz, H17) ; 6.92 (d, 3H, $^3J$ = 16.05 Hz, H7). $^{13}$C NMR (DMSO) : 141.7 ; 141.3 ; 141.1 ; 137.5 ; 128.9 ; 128.1 ; 127.6 ; 127.2 ; 125.7 ; 123.6 ; 122.8 ; 121.6 ; 121.1. Anal (calc) for C$_{42}$H$_{30}$S$_6$ : C 73.10 (73.57), H 6.78 (6.79), S 19.00 (19.64)

**Adsorption of compound 1 on Si(100).** Experiments were performed in an UHV system containing different chambers with base pressure less than 10$^{-10}$ Torr. Preparation of the silicon (100) surface and deposition of the molecules are done in two different chambers. The preparation of the Si(100) surface and the deposition process has been previously described.[11] The evaporation temperature of compound **1** is 305 ± 5°C, whereas the silicon substrate is held at room temperature during the evaporation. The duration of the evaporation is limited to a few seconds to get a submonolayer coverage of compound **1** on the clean and well-ordered silicon surface. Prior to the STM experiments, the W tips were electrochemically etched, cleaned in UHV and their radius of curvature was checked in field emission. All STM images were taken in constant current mode with negative sample voltages and a tunneling current of 60 pA.

**Monolayer of {C}$_{3h}$-2TV on HOPG.** To study 2D layers of {C}$_{3h}$-2TV oligomers, freshly cleaved surfaces of HOPG were used. Due to the small drift of the scanning tunneling microscope during an image acquisition in air when the scanning speed is low, the HOPG surface was always observed with the atomic resolution before the deposition of a drop to calibrate correctly the instrument. Nearly saturated solutions of {C}$_{3h}$-2TV oligomers in Cl$_2$CH$_2$ were then deposited onto the HOPG. The STM images were obtained with mechanically cut Pt/Ir tips at low sample voltages.

## Results

**Ab-initio calculations**: The contrast of an STM image depends on the variations of the topography and the electronic properties of the adsorbates. In the case of adsorbed monolayers on metallic surfaces, rarely the topographic factors dominate in the STM images. The contrast is mainly due to the electronic interaction between the tip and the molecule-surface system. As this contrast depends on the molecular levels and their coupling with the surface,[12] we have thus performed ab-initio calculations of the {C}$_{3h}$-2TV oligomer molecular orbitals. The aromatic and aliphatic moieties are different and other supramolecular structures formed with oligothiophenes have shown that the brightest area of an adsorbate, observed by STM, corresponds to the aromatic skeleton.[13,14] The alkyl chains are generally not observed or appear darker than the oligothiophene π-system. As a result, only the skeleton of the molecule, without the hexyl chains, was taken into account to model the chemical structure of the oligomer.

The calculations are treated with the local density approximation (LDA). For the computation, we used the DMOL code and the results are based on the spin-density functional of Vosko et al (VWN), as described elsewhere.[15] To build the molecule, we first optimize the dithienylethylene (DTE) branches and the benzene ring in LDA. The optimization of the whole system shows that the geometry of the molecule is planar and belongs to the $C_{3h}$ symmetry group. The separation between the highest occupied (HOMO) and the lowest unoccupied (LUMO) molecular orbitals is found to be 1.50 eV. As the Fermi level of the metallic surface is generally positioned within the HOMO-LUMO gap, the HOMO and LUMO levels will contribute predominantly to the STM image contrast at low polarization of the surface. The electronic structure of the highest occupied molecular orbital (HOMO) are thus shown in figure 2(a) and (b). The HOMO level is twofold degenerated and derives as expected from the interaction between the π orbitals along the whole molecule. Similarly to the HOMO level, the lowest unoccupied molecular orbital (LUMO) shown in 2(c) and (d) is also twofold degenerated. From these theoretical results, we can conclude that both levels are well delocalized on the whole skeleton and therefore the ternary symmetry of the {C}$_{3h}$-2TV oligomer should be visible in the STM images.

As olefins and aromatic systems such as benzene molecules have been shown to chemisorb on the Si(100) surface via cycloaddition reactions,[16,17] we have also performed electronic structure calculations of the molecular orbitals, when compound **1** is chemisorbed through the benzene ring to the Si dimers. Several bonding configurations have been found experimentally after adsorption: a single dimer bound benzene, corresponding to a [4+2] cycloadduct and different bridge configurations involving two Si dimers, which corresponds to different [4+4] cycloadducts.[18] As at room temperature, a conversion from the single dimer bound benzene to the bridging configurations has been observed, we focus on the [4+4] products, involving the central benzene ring of the {C}$_{3h}$ π-conjugated system and two Si dimers. Due to limitation of the computational time, we treat the most symmetric configurations, among the possible [4+4] cycloadducts: the symmetric bridge and the tight bridge configurations. To model the surface, a $Si_{12}H_{16}$ cluster has been used. The cluster contains two dimers in its top surface and bonds to sub-surface silicon atoms are terminated with hydrogen atoms. All atomic positions in the cluster are relaxed during the geometry optimization until the gradient is less than $10^{-3}$ hartree/bohr and the displacement less than $10^{-3}$ bohr. Optimization of the chemisorbed product geometry was performed by first attaching the benzene ring substituted with three vinyl groups to the two Si dimers of the surface and then allowing the geometry to relax. The benzene ring was then connected to the (DTE) chains, which were kept in the optimized orientation found for the ethylenic linkages and the electronic structure of the whole system was calculated for the conformation of fig.1.

Figure 3 shows the highest occupied orbitals (HOs) of the whole system for the symmetric bridging configuration. Comparison of the HOMO of the attached and free molecule shows that chemisorption breaks the orbital symmetry. Only two carbon atoms of the benzene ring, located between the Si-dimers keep their sp$_2$ character and hence contribute to the HOMO level. As shown in Fig. 3(a) a bean shape electron density contour is localized between the two atoms and can be described as a weak π-type bond. The Si-C bonds also contribute to the HOMO, giving a electron density on the benzene ring which is similar to the one found for the a benzene molecule chemisorbed on the Si(100) surface in the symmetric bridging configuration.[18] Consequently, the HOMO level for the whole system results from the electronic coupling between the HOMO level of the attached benzene ring alone and the HOMO level of the DTE ligand.

As the two other conjugated branches are connected to sp$_3$ carbons, they are only slightly involved in the HOMO. Conversely, the HOMO$_{-1}$, shown in Fig. 3(b), is mainly localized on these conjugated branches, leading also to an asymmetric orbital. The energy difference between the HOMO and HOMO$_{-1}$ levels is 0.67 eV.

For the tight bridging configuration, the orbitals of the DTE branches are less coupled due to the lower degree of symmetry. We obtain thus three different levels. As the cycloaddition of the benzene ring leads to a stronger π-type bond between both unreacted C atoms, since they are now first neighbors, the coupling of this state with the HOMO level of the 2TV ligand gives rise to the formation of a level with a lower energy in comparison with the symmetric bridge configuration. As a result this level corresponds now to the HOMO$_{-1}$ level of the whole system. As to the HOMO level, the calculation predicts that it is localized on another DTE branch, the HOMO$_{-2}$ level being localized on the third DTE branch. The energy difference between the HOMO and HOMO$_{-1}$ levels and the HOMO and HOMO$_{-2}$ levels are respectively 0.16 eV and 0.24 eV.

From these calculations of the electron density for the HOs, a general conclusion can be drawn. Due to the coupling between the HOMO level of the attached benzene ring with the HOMO level of the 2TV ligand, the electron density is significantly modified above the benzene ring in contrast with the electron density of the free system, observed in fig 2. This observation is true for both investigated [4+4] cycloadducts and can be also extended to the twisted bridge configuration. Therefore, any chemisorbed {C}$_{3h}$ π-conjugated system attached through a [4+4] cycloaddition reaction to the Si dimers is likely to have its appearance on the benzene ring modified in the STM image.

**Adsorption on Si(100):** Figure 4 shows a STM image of the Si(100) (2x1) surface after adsorption of compound **1**. The rows of silicon dimers are clearly apparent and various types of adsorbates are observed onto the surface. The smallest objects such as feature (a), which correspond to 15% of the observed adsorbates, may result from the breaking of some molecules during the evaporation process. On the other hand larger adsorbates such as (b) involve aggregates of two or more molecules. The most salient feature (c) of fig. 4 is that more than 50% of the adsorbates clearly exhibit the three characteristic branches of compound **1**. Although different geometries are observed, all branches have comparable lengths. The resolution of the dimer rows allows the estimation of the chain length: $11.0 \pm 1.7$ Å, in excellent agreement with the 10.8 Å calculated value. This implies that the whole molecule can be inserted in a circle of *ca* 25 Å diameter. Finally, 22 % of the adsorbates, like feature (d), show a comparable size to the size of feature (c), but appear featureless in the STM images. In the rest of the discussion, only objects, which exhibit three distinct branches will be considered.

A closer examination of the images of these adsorbates shows that three main types of structures can be distinguished (Fig. 5). A first type of adsorbate clearly exhibits a regular propeller-like shape (a). Molecules having this shape have been observed with both right and left rotations. This geometry, which exactly corresponds to the image expected for the actual chemical structure of the molecule **1** shows that both the chemical structure and the initial conformation of the molecule can survive the sublimation process. In a second type of adsorbate (b), the angles between the branches are still of 120° and the $C_{3h}$ symmetry is still apparent. However, the propeller shape is no longer observed. Such a structure of the conjugated side chains could result from rotations around the singles bonds connected to the thiophene rings. Finally, for a third type of adsorbates (c), the $C_{3h}$ symmetry is broken while two of the conjugated branches become almost collinear. This conformation may correspond to a structure resulting from rotations around the single bonds connecting the conjugated branches to the central benzene ring (Fig. 5f). Although care should be observed in the assignments of the hypothetic chemical structures of the conformers 5b and 5c, the STM images in Fig. 5 clearly show that compound **1** exists in different conformations onto the Si(100) surface.

All the conformers are generally observed in different orientations on the surface. The central part of conformers 5a and 5c are positioned either on the top of the dimer rows or in between, without a significant change of the contrast. Conversely, the adsorption site of the central part of conformers 5b is found in the troughs between two dimer rows. In all cases, the branches have no particular orientation with respect to the dimer rows.

**Adsorption on HOPG:** Figure 6 shows an STM image of a layer of $\{C\}_{3h}$-2TV oligomers. A spontaneous ordering is observed with a honeycomb structure. Although the components of the structure are fuzzy, they seem to have a geometry similar to the geometry of molecule **1**. Their size is deduced from the HOPG surface, which was observed with the atomic resolution before the deposition of the oligomers, as shown in the inset of fig. 6. Between the adjacent bright central parts of the components, we measure an average length of 28 ± 3 Å. This size corresponds to the expected size of compound **1** and the components of the honeycomb structure are assigned to be the backbone of a $\{C\}_{3h}$-2TV oligomer. Such a structure for the backbone is also in agreement with our ab-initio calculations, since the HOMO and LUMO levels have been found to be fully delocalized on the oligomer backbone.

As we have identified the skeleton of the molecule and the hexyl chains are supposed to lie between the bright branches of the oligomers, the dark regions of the image contain the hexyl chains. Such a result is in agreement with previous STM studies of oligomers adsorbed on HOPG, where the bright contrast was dominated by the aromatic moieties and the dark one by the alkyl chains.[19] Taking into account the steric volume of the hexyl chains, we can therefore model supramolecular structures. The one, which fits the best to the molecular arrangement, is shown in figure 7. In this case the packing structure is given by the lattice vectors **A** and **B**, with one molecule per unit cell. The molecular lattice has the same symmetry as the one for the HOPG surface. We find that **A** =10 x **a** and **B** =10 x**b**, with an angle between the lattice vectors **A** and **B** of 120°, where **a** and **b** are the lattice vectors of the HOPG surface. The area per molecule is thus 524 Å$^2$.

As previously observed for other supramolecular structure,[20] the {C}$_3$-2TV oligomers cover the surface as densely as possible. Although it is difficult for the hexyl chains to interdigitate, the steric volume occupied by the hexyl chains is yet small enough so that they can interact. In the proposed structure of figure 7, a few hexyl chains slightly overlap and small rearrangements of the alkyls chains probably occur to avoid the overlapping. As attempts to image the molecule **1**, which is not substituted by alkyl chains, have been made on HOPG, but were not successful, we believe that the ordering is caused by the interaction between the alkyl chains. It is generally observed that alkanes and alkyl chains of linear oligothiophenes are oriented parallel to one of the crystal axes of HOPG.[13,21] In the case of the proposed structure for the {C}$_{3h}$-2TV molecules, the hexyl chains have three possible orientations, which are deduced from each other by a rotation of 60°. Although the orientation of the lattice vectors a and b of the HOPG surface is not known when we image the molecules and thus may differ from the one shown in the inset of fig.6, the symmetry observed for the hexyl chains could be related to the orientation of the HOPG lattice vector. Even though a small rearrangement is necessary to avoid the overlapping of a few hexyl chains, the alignment of most of the hexyl chains with the crystal axes could lead to the formation of a molecular arrangement with a three fold symmetry as the one observed in fig. 6.

A closer look to the STM image reveals that the central core appears slightly brighter than the rest of the oligothiophene system. This part of the oligomer is made up of a benzene ring whereas the rest of the molecule consists of alternative thiophene rings and C=C bonds. Furthermore the benzene ring does not support any alkyl chains. Only the thiophene rings are substituted by the hexyl chains. As a result, the benzenic core of the oligomers may interact in a different manner with the HOPG surface in comparison with the three branches. This interaction may be at the origin of the orientation of the 2D layer with respect to the HOPG surface.

## Discussion

The results obtained on the Si(100) and on the HOPG surfaces contrast clearly. The attachment of alkyl chains to the molecule **1** allows the formation of a 2D layer on HOPG, whereas the molecules **1**, vapour deposited on the Si(100) surface at a small coverage, interact with the surface in a way that they cannot diffuse on the surface to form an ordered layer. As the interaction between the molecules **1** and the Si(100) surface is strong enough to immobilize the molecules onto the surface, the observation of individual oligomers on the Si(100) surface is made possible at room temperature. Therefore this surface forms an interesting support to characterize the structure of oligomers after their synthesis.

However, some cares must be taken in order to interpret the STM images obtained on the Si(100) surface and then determine the oligomer structure. Firstly, the oligomers show different conformations on the Si(100) surface, whereas a single conformation seems to be observed when the molecules are substituted with hexyl chains and deposited on HOPG. This result suggests that conformational changes involving rotations around single bonds occur during the sublimation process. In the frame of this hypothesis, adsorption onto the silicon surface would freeze the various conformations thermally produced during the sublimation process. In other words, the observed STM images would give a representation of the population of the various conformers present in the gaz phase before adsorption onto the Si surface. Such different conformers would not exist after the chemical synthesis of the oligomers.

A second point to discuss is the reactivity of the Si(100) surface. Indeed, it is known that the reaction of unsaturated organic molecules with the Si(100) dimers is generally facile.[22,23] Such reaction generally lead to new chemical products. The chemical structure of the final product can be thus quite different from the structure of the organic compound before the reaction. If we focus on the contrast variation of compound **1** adsorbed on the Si(100) surface and showing the three ligands, we generally do not observe a significant contrast variation from one ligand to another, but we can see two types of contrast in the central part of the molecules. Such an example is shown in fig. 8. A comparison of the height profiles between the molecule in fig. 8(a) and the clockwise screw propeller molecule in fig. 8(b) gives a difference of height of 1.1 Å. Although, we cannot neglect the possibility to have defect sites on the surface below those bright molecules, what could modify the contrast variation, a variation of the brightness can also be attributed to the reaction of the molecule with the Si dimers. Indeed, the comparison of the theroretical occupied molecular orbitals of the free molecule and the molecule after reaction of the benzene ring with the Si surface has shown an increase of the electronic density on the benzene ring after reaction. A calculation of the tunneling current is however needed to reach a definitive conclusion.

Nevertheless, in the case of compound **1**, the mechanisms of the reaction are likely to be different from the ones observed for simple alkenes. Indeed, due to the size of the molecule, steric hindrance might prevent the formation of short lived intermediates, which seem to be necessary to produce a reaction with the Si dimers, as it was shown for ethylene.[22] As the adsorption of compound **1** on the Si(100) surface shows different conformers lying on the surface in a large number of orientations with respect to the Si dimers, the constitutive blocks of compound 1 with the highest symmetry is more likely to react with the surface. The benzenic core has a higher degree of symmetry than the thiophene rings or the ethylene linkages and is more favoured to react. Such a hypothesis would explain why the STM observation of the DTE branches on the Si(100) surface do not show any significant contrast variation from adsorbed molecule to another. Therefore, the number of bonds, which are cleaved to make covalent bonds with Si atoms on the surface, is expected to be quite small and will not strongly alter the chemical structure of the oligomer. This result agrees with our observation, since the majority of the adsorbates show the three ligands, and make the use of the Si(100) surface relevant to observe isolated complex organic molecules at room temperature.

## CONCLUSION

Conjugated oligomers with a ternary symmetry have been synthesized. By substituting the thiophene ring of the oligomers with alkyl chains, the oligomers form an ordered 2D layer, when they are deposited on a HOPG surface from a solution. Because of the geometry of the molecule, the ordering differs from the one obtained with linear oligothiophene system, where the alkyl chains can interdigitate.

While the literature shows numerous examples of oligomers forming 2D layers on HOPG or $MoS_2$ surfaces, adsorption of the same oligomers on the Si(100) surface has never been done. Our results show that high resolution images of individual oligomers are possible when the molecules are adsorbed onto the Si(100) surface in UHV, what is rarely the case at the liquid-solid interface of a HOPG substrate, due to the imaging conditions. As the diffusion of the oligomers does not occur on the Si(100) surface at room temperature, this study reveals that semiconductor surfaces such as the Si(100) surface may be of interest to characterize new synthesized oligomers in an individual manner.

*Email address: grandidier@isen.iemn.univ-lille1.fr

(1) Rabe, J.P.; Buchholz, S. *Science* **1991**, *253*, 424.

(2) Mc Gonigal, G.C.; Bernhardt, R.H.; Thomson, D.J. *Appl. Phys. Lett.* **1990**, *57*, 28.

(3) Krömer, J.; Rios-Carreras, I.; Furhmann, G.; Musch, C.; Wunderlin, M.; Debaerdemaecker, T.; Mena-Osteritz, E.; Bäuerle, P. *Angew. Chem. Int. Ed.* **2000**, *39*, 3481.

(4) Böhringer, M. ; Schneider, W-D.; Berndt, R. *Surf. Sci.* **1998**, *408*, 72.

(5) Böhringer, M. ; Morgenstern, K.; Schneider, W-D.; Wühn, M.; Wöll, C.; Berndt, R. *Surf. Sci.* **2000**, *144*, 199.

(6) Furukawa, M. ; Tanaka, H.; Kawai, T. *J.Chem. Phys.* **2001**, *115*, 3419.

(7) Elandaloussi, H. ; Frère, P. ; Richomme, P. ; Orduna, J. ; Garin, J. ; Roncali, J. *J. Am. Chem. Soc.* **1997**, *119*, 10774.

(8) Roncali, J. *Acc. Chem. Res.* **2000**, *33*, 147.

(9) Jestin, I. ; Frère, P. ; Mercier, N. ; Levillain, E. ; Stiévenard, D. ;Roncali, J. *J. Am. Chem. Soc.* **1998**, *120*, 8150.

(10) Elandaloussi, E. ; Frère, P. ; Roncali, J. *Chem. Commun.* **1997**, 301.

(11) Grandidier, B. ; Nys, J.P. ; Stiévenard, D. ; Krzeminski, C. ; Delerue, C. ; Frère, P. ; Blanchard,P. ; Roncali, J. *Surf. S*ci. **2001**, *473*, 1.

(12) Magoga, M.; Joachim, C.*; Phys. Rev. B* **1997**, *56*, 4722.

(13) Bäuerle, P.; Fisher, T.; Bidlingmeier, B.; Stabel, A.; Rabe, J.P. *Angew. Chem. Intl. Ed. Engl.* **1995**, *34*, 303.

(14) Azumi, R. ; Götz, G. ; Bauerle, P. *Synt. Met.* **1994**, *101*, 569.

(15) Krzeminski, C.; Delerue, C.; Allan, G. ; Haguet, V. ; Stiévenard, D. ; Frère, P. ; Levillain, E. ; Roncali, J. *J. Chem. Phys.* **1999,** *111*, 6643.

(16) Lopinski, G.P.; Moffatt, D.J.; Wolkow, R.A. *Chem. Phys. Lett*. **1998**, *282*, 305.

(17) Kong, M.J.; Teplyakov, A.V.; Lyubovitsky, J.G.; Bent, S.F. *Surf. Sci.* **1988**, *411*, 286.

(18) R.A. Wolkow, G.P. Lopinski, D.J. Mofatt, Surf. Sci. **1998**, 416, L1107.

(19) Claypool, C.L.; Faglioni, F.; Goddard, W.A.; Gray, H.B.; Lewis, N.S.; Marcus, R.A.; *J.phys. Chem. B* **1997**, *101*, 5978.

(20) Azumi, R.; Götz, G.; Debaerdemaeker, T.; Bäuerle, P. *Chem. Eur. J.* **2000**, *6*, 735.

(21) Cyr, D.M.; Venkataramann, B.; Flynn, G.W., *Chem. Mater*. **1996**, *8*, 1600.

(22) Liu, H.; Hamers, R.J. *J. Am. Chem. Soc.* **1997**, *119*, 7593.

(23) Hamers, R.J.; Hovis, J.S.; Greenlief, C.M.; Padowitz, D.F.; *Jpn. J. Appl. Phys.* **1999**, *38*, 3879.

**Figure 1.** Synthesis of a {C}$_{3h}$-oligomer **1** involving three dithienylethylene branches.

**Figure 2.** a) and b) HOMO of the {C}$_{3h}$-2TV oligomer without the hexyl chains. c) and d) LUMO of the {C}$_{3h}$-2TV oligomer without the hexyl chains. Both levels are degenerated twofold.

**Figure 3.** Top view of the compound **1** chemisorbed on the Si(100) surface in the symmetric bridge configuration showing the calculated HOMO (a) and HOMO$_{-1}$ (b).

**Figure 4.** STM image of the Si(100) surface after the deposition of the compound 1. The image was acquired with a sample bias of –2.9 V and a tunneling current of 60 pA. The different types of features, (a), (b), (c) and (d), observed on the surface are described in the text. The image size is 174 × 214 Å$^2$. The grey scale ranges from 0 (black) to 6.1 Å (white).

**Figure 5.** STM images of different types of adsorbates, which show the three ligands, and their associated chemical structure. The image was acquired with a sample bias of –2.9 V. The image size is 62 × 62 Å$^2$. The grey scale ranges from 0 (black) to 4.2 Å (white) for the three STM images.

**Figure 6.** Constant current STM image of {C}$_3$-2TV oligomers adsorbed on HOPG. The image was acquired with a sample voltage of +200 mV and a tunneling current of 300 pA. The inset shows the HOPG surface with the atomic resolution.

**Figure 7.** STM image of {C}$_3$-2TV oligomers with the model of the molecular arrangement.

**Figure 8.** STM images molecules 1 showing in (a) their brightest part positioned above the central core and in (b) a similar contrast variation between the central core and the branches. For both STM images, the size is $62 \times 62$ Å$^2$. The height profiles along the directions indicated by the black arrows are given to compare the contrast variation.

Figure 1

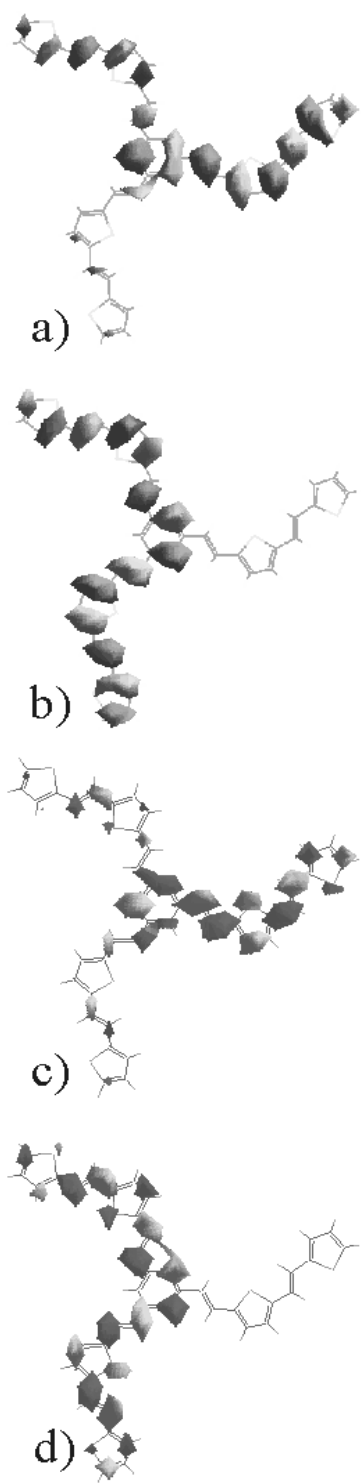

Figure 2

a)

b)

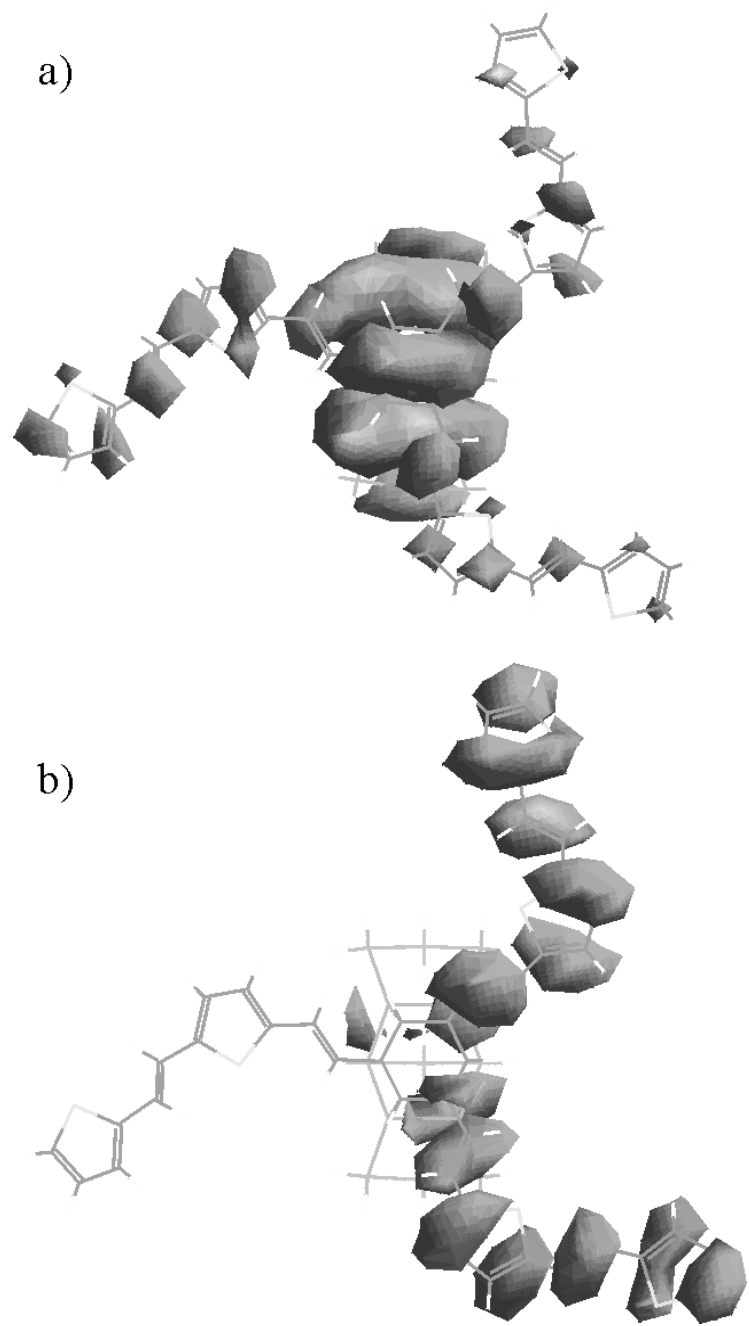

Figure 3

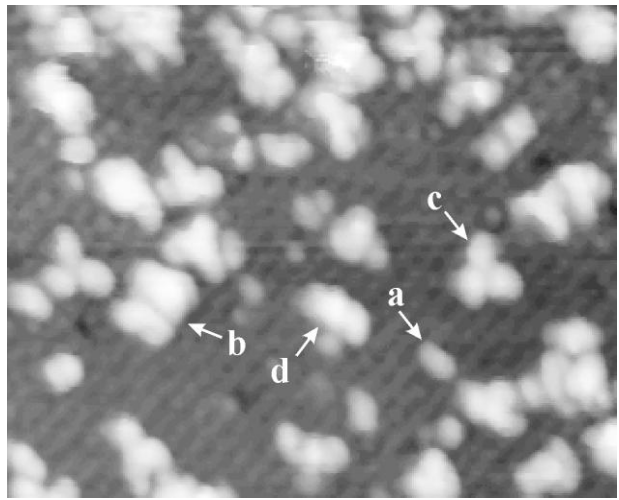

Figure 4

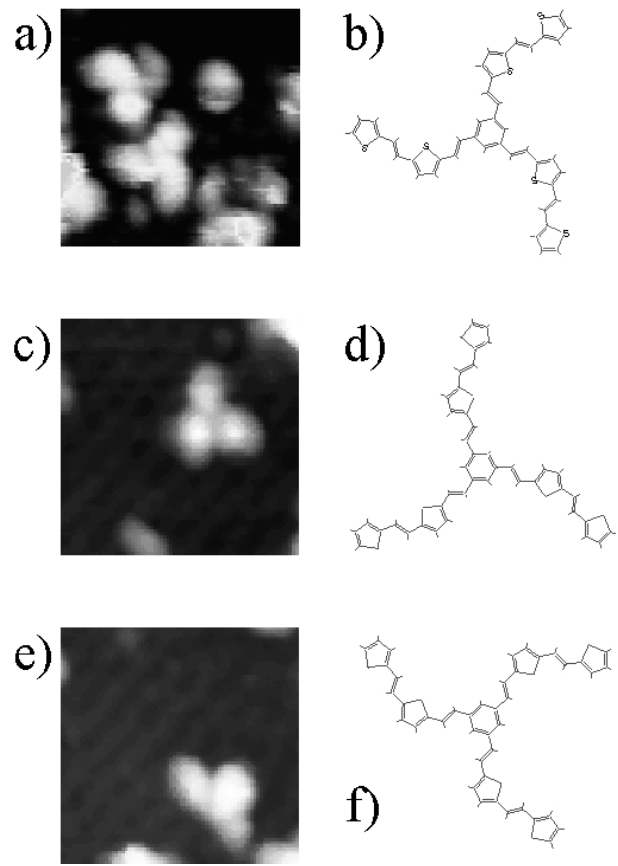

Figure 5

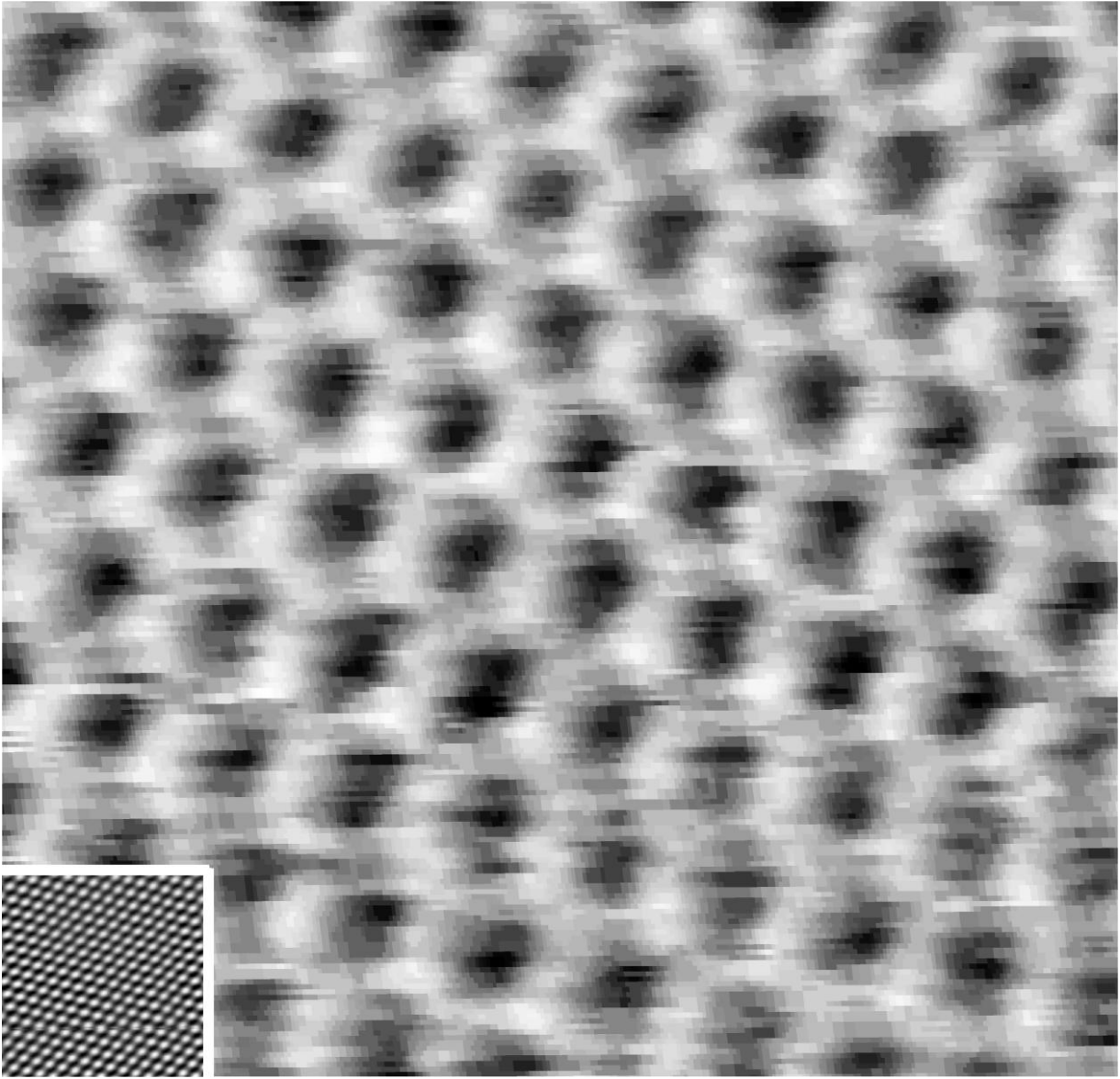

Figure 6

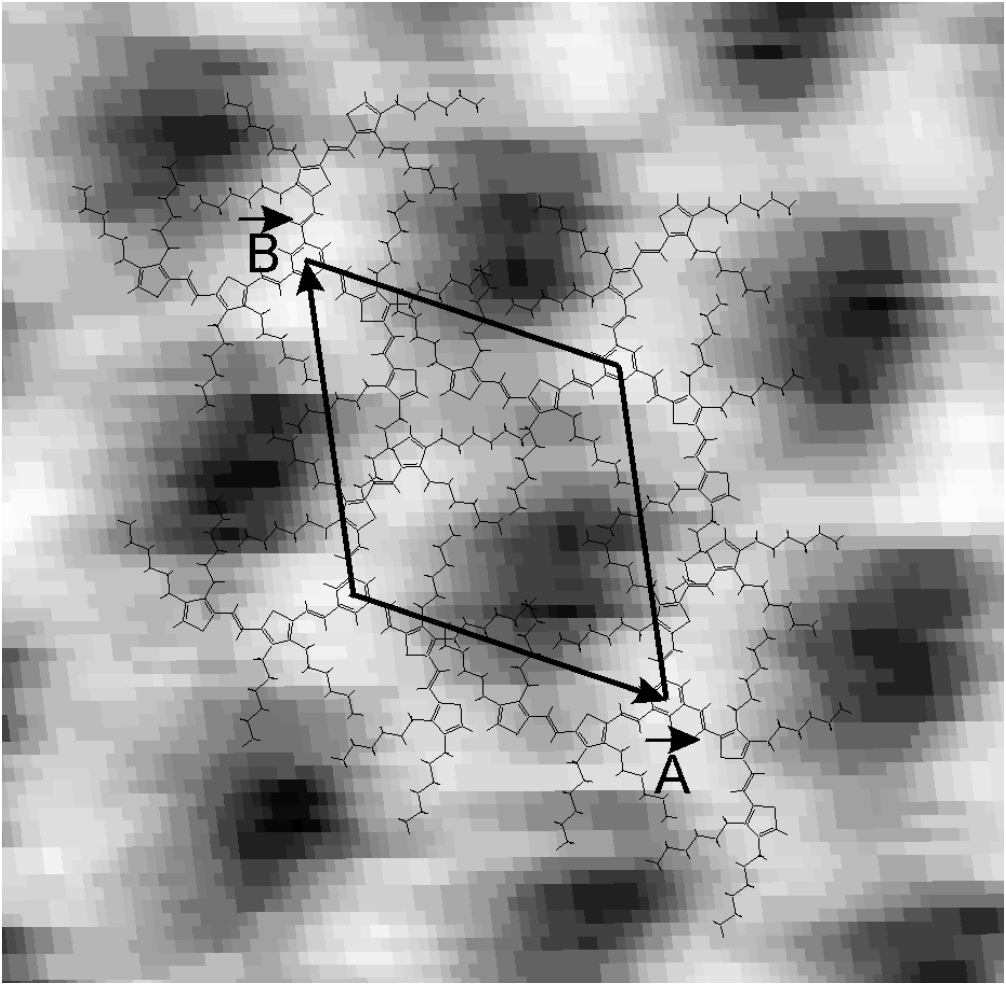

Figure 7

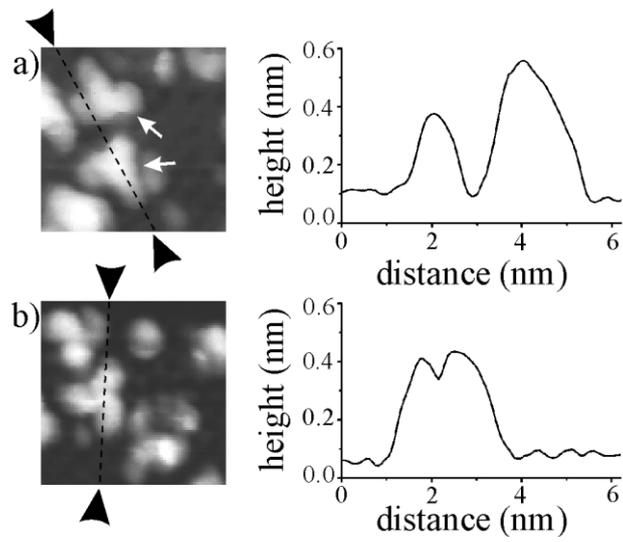

Figure 8